\documentclass[smallcondensed]{svjour3}    % onecolumn (standard format)

\smartqed  % flush right qed marks, e.g. at end of proof
\usepackage{graphicx, subfigure}
\usepackage{amssymb,amsmath}
\usepackage{natbib}
\usepackage{txfonts}
\usepackage{url}
\usepackage{pifont}
\usepackage{multirow}% http://ctan.org/pkg/multirow
\usepackage{hhline}% http://ctan.org/pkg/hhline

\newcommand{\Msun}{M_\odot}
\newcommand{\Rsun}{R_\odot}

\newcommand{\Ms}{M_{\star}}
\newcommand{\Rs}{R_{\star}}
\newcommand{\Ps}{P_{\star}}

\newcommand{\Os}{\Omega_{\star}}

\newcommand{\Is}{I_\star}

\newcommand{\wsat}{\omega_{\rm sat}}

\newcommand{\dd}{\mathrm{d}}

\usepackage[normalem]{ulem}
\usepackage{color}
\definecolor{blue}{RGB}{0,0,255}
\definecolor{red}{RGB}{255,0,0}
\definecolor{green}{RGB}{0,200,0}
\definecolor{black}{RGB}{0,0,0}

\begin{document}

\title{Correction to: Effect of the rotation and tidal dissipation history of stars on the evolution of close-in planets}
\subtitle{Correction to: Celest Mech Dyn Astr (2016) 126:275Ð296
\url{https://doi.org/10.1007/s10569-016-9690-3}}

%\titlerunning{Short form of title} % if too long for running head

\author{Emeline Bolmont$^{1, 2, 3}$         \and
        St\'ephane Mathis$^2$ %etc.
}


\institute{Emeline Bolmont \\
              \email{emeline.bolmont@unige.ch}\\
$^1$ NaXys, Department of Mathematics, University of Namur, 8 Rempart de la Vierge, 5000 Namur,
Belgium \\
$^2$ AIM, CEA, CNRS, Universit\'e Paris-Saclay, Universit\'e Paris Diderot, Sorbonne Paris Cit\'e, F-91191 Gif-sur-Yvette, France\\
$^3$ Observatoire Astronomique de l'Universit\'e de Gen\`eve, Universit\'e de Gen\`eve, Chemin des Maillettes 51,
1290 Versoix, Switzerland}        
\maketitle

\begin{abstract}
This is an erratum for the publication \citealt{Bolmont16} (Celestial Mechanics and Dynamical Astronomy, 	126, 275-296). 
There was a small mistake for the spin integration of our code which we corrected and we take advantage of this erratum to investigate a bit further the influence of a planet on the spin of its host star. 

%% \PACS{PACS code1 \and PACS code2 \and more}

\end{abstract}

The code used in \citet{Bolmont16} had a small mistake for the spin integration. 
While this does not change the results of the simulations and most of the plots, the correction of this bug has an impact on Figure~5.

Figure~5 displayed the difference between the rotation period of a star with an outward-migrating planet and the rotation of the same star without planet. 
This Figure has the purpose of showing the impact of the planet on the rotation of its host star.
The variations of the stellar rotation period induced by the migration of the planet are quite small compared to the spin-up and spin-down amplitude, plotting the difference allows to magnify and isolate this effect.

We take advantage of this erratum where we corrected our code to explain and investigate the evolution of the rotation period of a star with a migrating planet compared to the evolution of the rotation period of a star without planet.
We define here $\delta P$ as the difference between the rotation period of the star with planet $P_{\star, {\rm pl}}$ and the rotation period of the star without planet $P_{\star, {\rm no~pl}}$
\begin{equation}
\delta P(t) = P_{\star, {\rm pl}}(t) - P_{\star, {\rm no~pl}}(t).
\end{equation}
Similarly we define $\delta \Omega = \Omega_{\star, {\rm pl}}(t) - \Omega_{\star, {\rm no~pl}}(t)$ as the difference between the spin of the star with planet and the spin of the star without planet.
$\Omega_\star$ is here the angular velocity of the star: $\Omega_\star = 2\pi/P_{\star}$.

The star is considered here to have a solid rotation. 
The planets are on coplanar and circular orbits.
We remind here the Equations governing the evolution of the spin $\Os$ of the star (Equation~14 of \citealt{Bolmont16})
\begin{align} \label{truc}
\begin{split}
\frac{\dd \Is\Os}{\dd t} &= -K\Omega_\star^\mu \omega_{\rm sat}^{3-\mu}\left(\frac{R_{\star}}{\Rsun}\right)^{1/2}\left(\frac{M_{\star}}{\Msun}\right)^{-1/2} \\
 & + \frac{h}{2T_\star}\left[1-\frac{\Omega_{\star}}{n}\right], 
\end{split}
\end{align}
where $\Is$ is the stellar moment of inertia which varies as the radius of the star $\Rs$ evolves \citep[we use the evolution grids of][]{Siess2000}, $h$ the orbital angular momentum, $n$ the mean orbital angular frequency, $T_\star$ is the stellar dissipation timescale (see Eq. 13 of \citealt{Bolmont16}, note that $1/T_\star \propto a^{-8}$).
$K$, and $\wsat$ are parameters of the stellar wind model from \citet{Bouvier1997}, which is valid for Main Sequence stars in the studied mass range. 
If  $\Omega_\star>\wsat$, $\mu = 1$ and if $\Omega_\star<\wsat$, $\mu = 3$.
Equation~\ref{truc} becomes
\begin{align} \label{truc2}
\begin{split}
\frac{\dd\Os}{\dd t} = \frac{1}{\Is}\left[-\dot{I}_\star \Os -K\Omega_\star^\mu \omega_{\rm sat}^{3-\mu}\left(\frac{R_{\star}}{\Rsun}\right)^{1/2}\left(\frac{M_{\star}}{\Msun}\right)^{-1/2} + \frac{h}{2T_\star}\left[1-\frac{\Omega_{\star}}{n}\right]\right]. 
\end{split}
\end{align}
If we write this equation in terms of the stellar rotation period $\Ps = 2\pi/\Os$, we get
\begin{align} \label{dPstar}
\begin{split}
\frac{\dd\Ps}{\dd t} = \frac{-2\pi}{\Is}\left[-\dot{I}_\star \Os^{-1} -K\Omega_\star^{\mu-2} \omega_{\rm sat}^{3-\mu}\left(\frac{R_{\star}}{\Rsun}\right)^{1/2}\left(\frac{M_{\star}}{\Msun}\right)^{-1/2} + \frac{h}{2T_\star\Os^2}\left[1-\frac{\Omega_{\star}}{n}\right]\right]. 
\end{split}
\end{align}

\begin{figure*}[htbp!]
	\centering
%        \begin{center}
        \includegraphics[width=\linewidth]{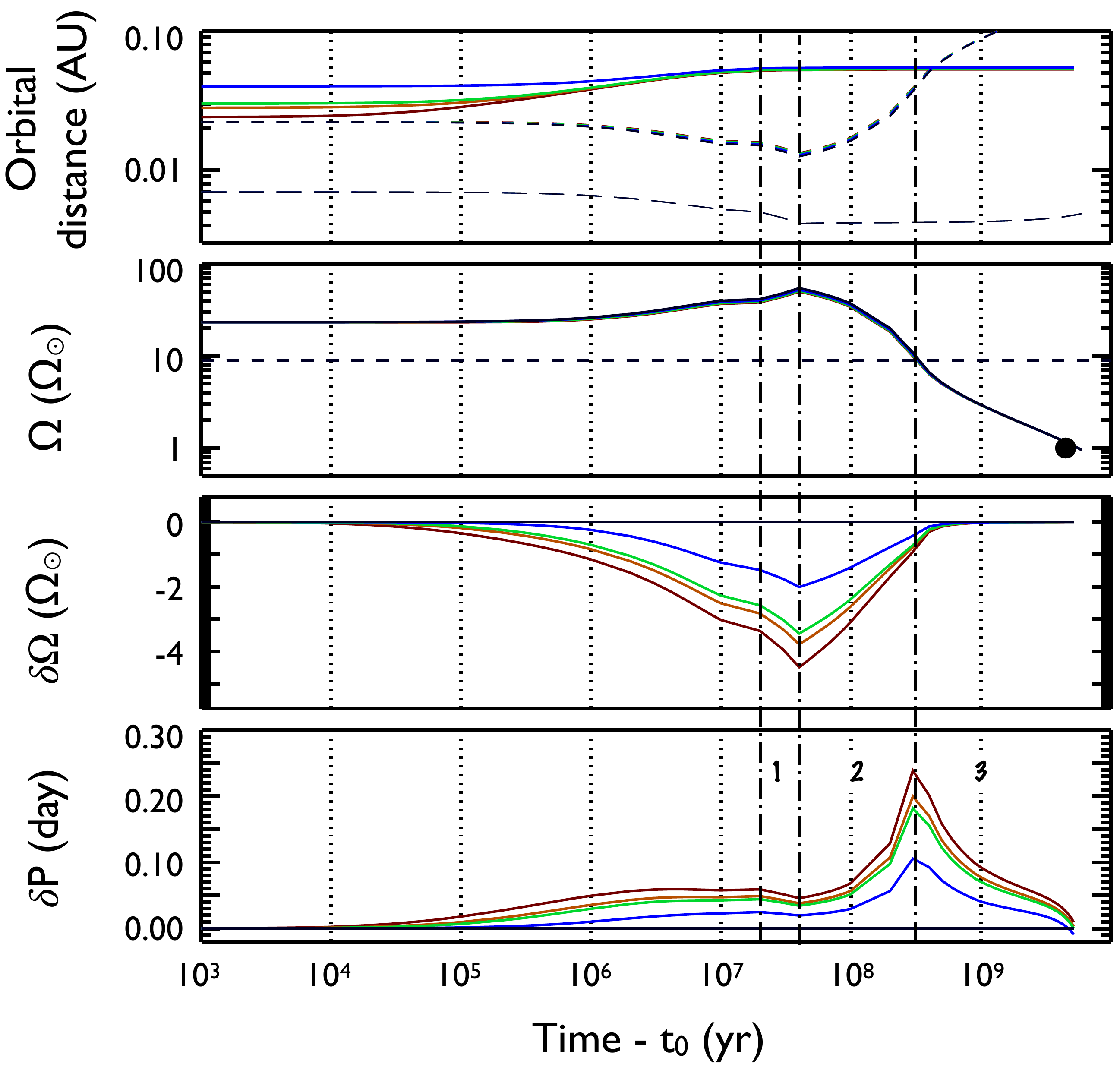}
        \caption{Tidal evolution of Jupiter mass planets around an initially fast rotating Sun-like star. Top panel: evolution of the semi-major axis of the planet (colored full lines), corotation distance (dashed lines) and stellar radius (long dashed lines). Middle top panel: evolution of the spin of the star (full lines) and the saturation spin $\wsat$ (dashed line). The black lines correspond to the case with no planet and the present day rotation of the Sun is represented by the black dot. Middle bottom panel: evolution of $\delta \Omega$. Bottom panel: evolution of $\delta P$. Note that $\delta \Omega$ for the closest planet (in blue) also changes sign when $\delta P$ does towards the end of the evolution, however the scale allowing us to display the full evolution of $\delta \Omega$ does not allow to show this sign reversal.}
        \label{a_spin_dP}
%        \end{center}
        \end{figure*}

%Let us consider two cases: an initially fast rotating star ($P_{\star, 0} = 1.2$~day) and an initially slow rotating star ($P_{\star, 0} = 8.0$~day).
%Planets around the fast rotating star migrate outwards and therefore acts to slow down the rotation of the star. 
%Planets around the slow rotating star migrate inwards and therefore acts to spin up the rotation of the star.
Let us consider an initially fast rotating Sun-like star ($\Ms = 1~\Msun$, $P_{\star, 0} = 1.2$~day). 
Figure~\ref{a_spin_dP} shows the orbital evolution of Jupiter mass planets around a Sun-like star and the evolution of the spin of the star.
As in \citet{Bolmont16}, we consider that the initial time of our simulations is 5~Myr. 
The planets are distributed between an initial semi-major axis of 0.022~AU and 0.04~AU.
The planets, initially outside the corotation distance, migrate outwards and therefore by conservation of angular momentum act to slow down the contraction induced spin-up of the star.
$\delta \Omega$ has a straightforward evolution, the star spins down when the planet is migrating outwards. 
The closer the planet to the corotation distance, the farther the migration and the larger the spin-down of the star.
Once the migration has stopped the winds tend to spin down faster the star which rotates faster (the star without planet), and slower the star which rotates slower (the star with planet).
This leads to a convergence towards 0 of $\delta \Omega$.

However, the evolution of $\delta P$ is not as straightforward. 
It increases as it should until an age of $\sim10$~Myr, then decreases until the end of the contraction phase (where the stellar rotation is the fastest), increases significantly until $\sim 500$~Myr (when the star passes from a saturated regime to an unsaturated one) and then decreases towards 0.

To explain this behavior, let us consider two stars 1 and 2 of same age but different rotation periods, $P_1 = 2\pi/\Omega_1$ and $P_2 = 2\pi/\Omega_2$.  
As the stars have the same age, they have the same radius $R_1 = R_2 = \Rs$ and moment of inertia $I_1 = I_2 = \Is$ (here, we neglect the impact of rotation on stellar structure and evolution; e.g. \citealt{2009pfer.book.....M} and references therein).
From Equ.~\ref{dPstar}, we can compute the derivative of $\Delta P = P_1-P_2$
\begin{equation}
\frac{\dd \Delta P}{\dd t} = \frac{2\pi}{\Is}\frac{\Omega_2 - \Omega_1}{\Omega_1\Omega_2}\times
\begin{cases}
\displaystyle{\dot{I}_\star - K\Omega_1\Omega_2\left(\frac{R_{\star}}{\Rsun}\right)^{1/2}\left(\frac{M_{\star}}{\Msun}\right)^{-1/2}}, & \text{if }\{\Omega_1, \Omega_2\}<\wsat, \\
\displaystyle{\dot{I}_\star + K\wsat^2\left(\frac{R_{\star}}{\Rsun}\right)^{1/2}\left(\frac{M_{\star}}{\Msun}\right)^{-1/2}}, & \text{if }\{\Omega_1, \Omega_2\}>\wsat.
\end{cases}
\end{equation}

Let us consider that star 1 spins slower than star 2 (e.g., star 1 had an outward-migrating planet, star 2 has no planet) so that $\Omega_1 < \Omega_2$.
If the two stars are in the saturated regime, then $\displaystyle{\frac{\dd \Delta P}{\dd t}}$ has the sign of $\dot{I}_\star + K\wsat^2\displaystyle{\left(\frac{R_{\star}}{\Rsun}\right)^{1/2}\left(\frac{M_{\star}}{\Msun}\right)^{-1/2}}$.
This means that there is a critical derivative of the moment of inertia $\dot{I}_{\rm crit}$ defined as
\begin{equation}
\dot{I}_{\rm crit} = -K\wsat^2\displaystyle{\left(\frac{R_{\star}}{\Rsun}\right)^{1/2}\left(\frac{M_{\star}}{\Msun}\right)^{-1/2}}.
\end{equation}
If the two stars are in the unsaturated regime, then $\displaystyle{\frac{\dd \Delta P}{\dd t}}$ has the sign of $\dot{I}_\star - K\Omega_1\Omega_2\displaystyle{\left(\frac{R_{\star}}{\Rsun}\right)^{1/2}\left(\frac{M_{\star}}{\Msun}\right)^{-1/2}}$.
In our example, the unsaturated regime is reached when the star is on the main sequence so when $\dot{I}_\star \approx 0$, so that here $\displaystyle{\frac{\dd \Delta P}{\dd t}}$ is negative.
We can therefore identify the different phases of the evolution of $\delta P = P_1-P_2$.
\begin{enumerate}
\item[-] If the stars are in the saturated regime and if $\dot{\Is} < \dot{I}_{\rm crit}$, which is verified if the star experiences a strong contraction (i.e. just before the main sequence, or in absolute values: $|\dot{\Is}| > |\dot{I}_{\rm crit}|$), $\delta P$ decreases (the rotations converge): this is the phase number 1 of Fig.~\ref{a_spin_dP}.
\item[-] If the stars are in the saturated regime and if $\dot{\Is} > \dot{I}_{\rm crit}$, which is verified if the stellar moment of inertia does not evolve much (i.e. during most of the PMS and the MS), $\delta P$ increases (the rotations diverge). 
This is what happens from the end of the contraction phase ($\sim50$~Myr) to the unsaturated/saturated transition ($\sim 500$~Myr): this is the phase number 2 of Fig.~\ref{a_spin_dP}.
\item[-] If the stars are in the unsaturated regime and on the main sequence, $\delta P$ decreases (the rotations converge): this is the phase number 3 of Fig.~\ref{a_spin_dP}.
\item[-] If the stars are in the unsaturated regime and at the beginning of the giant branch (where the radius of the star increases sufficiently so that $\dot{I}_\star > K\Omega_1\Omega_2\displaystyle{\left(\frac{R_{\star}}{\Rsun}\right)^{1/2}\left(\frac{M_{\star}}{\Msun}\right)^{-1/2}}$, $\delta P$ increases (the rotations diverge).
We can see this effect starting to operate as the slope of $\delta P$ changes its inflection around 3~Gyr (bottom panel of Fig.~\ref{a_spin_dP}).
But this trend cannot really be seen because at late ages, the planets begin to fall towards the star making it accelerate sightly. 
\end{enumerate}

\begin{figure*}[ht!]
     \begin{center}
        \subfigure[$\Ms = 0.6~\Msun$]{
            \label{fig:06}
            \includegraphics[angle=0,width=0.48\textwidth]{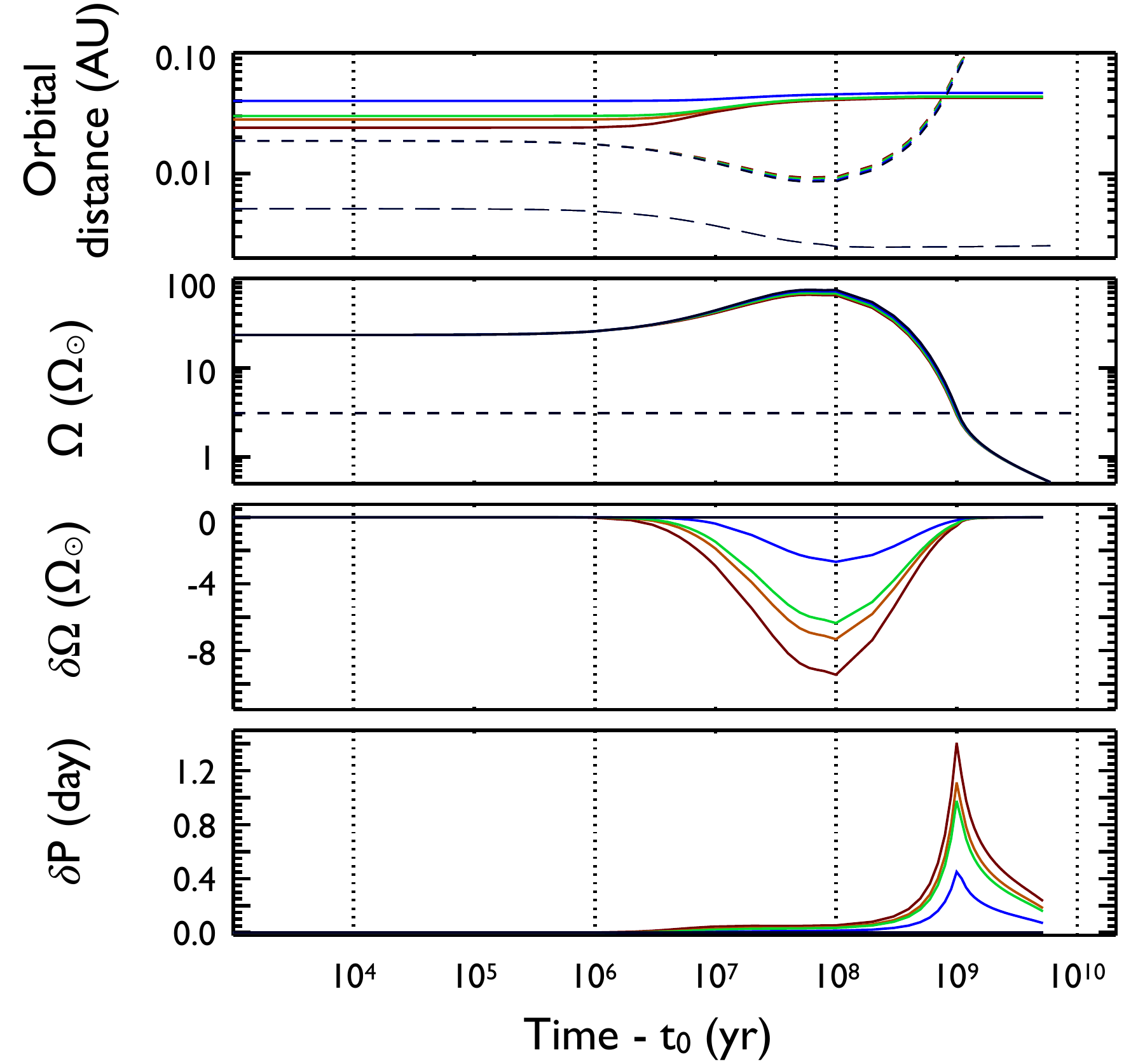}
        } \hspace{0.0cm}
        \subfigure[$\Ms = 1.2 ~M_{\odot}$]{
           \label{fig:12}
           \includegraphics[angle=0,width=0.48\textwidth]{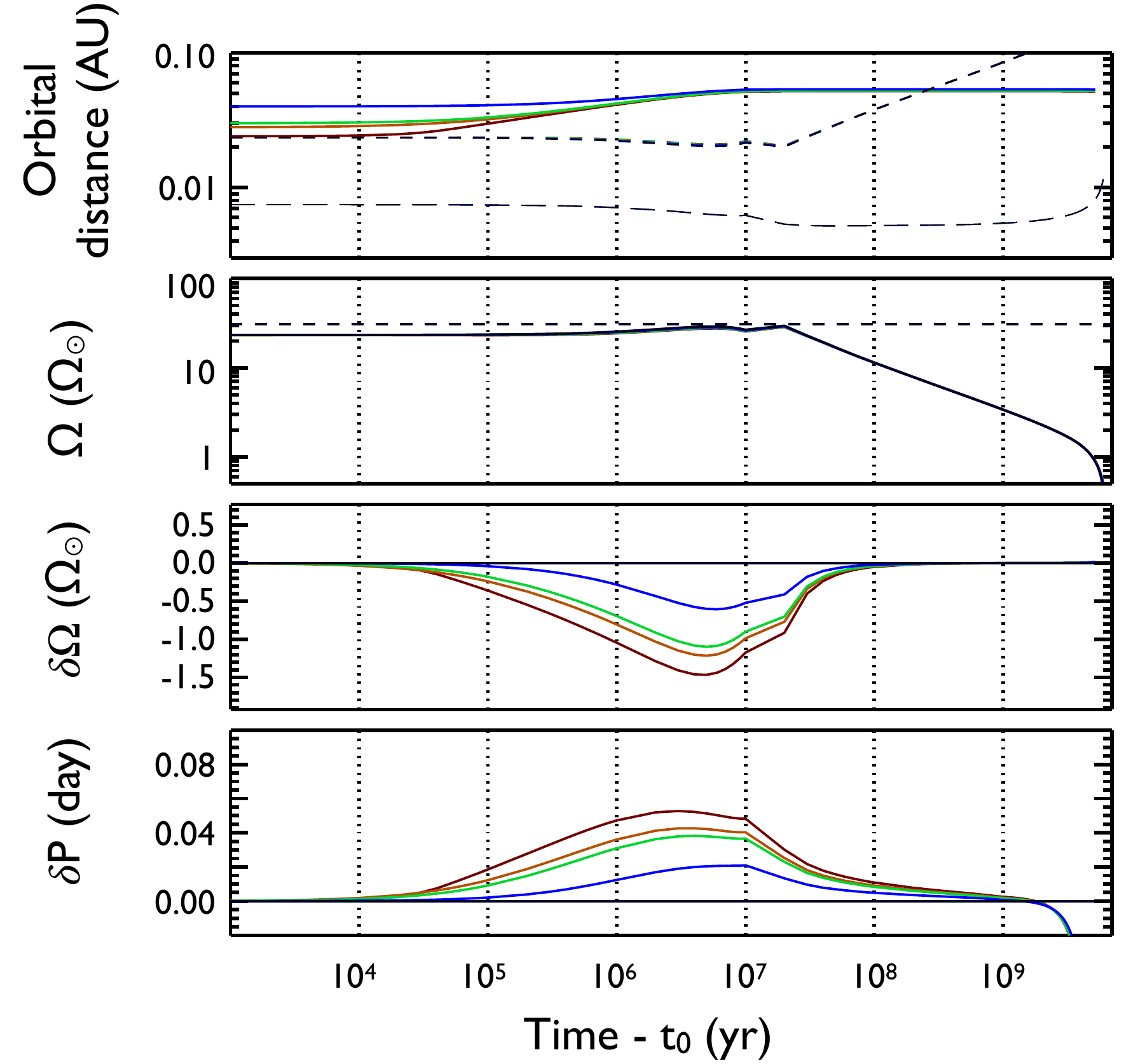} 
        }
    \end{center}
    \caption{Same as Fig.~\ref{a_spin_dP} but for different masses: (a) $0.6~\Msun$, (b) $1.2~\Msun$. Note that the scales are different for the two cases.}
   \label{a_spin_dP_Ms}
\end{figure*}

Figure~\ref{a_spin_dP_Ms} shows the evolution of Jupiter mass planets around stars of different masses, $\Ms = 0.6~\Msun$ and $\Ms = 1.2~\Msun$.
As expected, the effect of the planet on the stellar rotation decreases as stellar mass increases.
For $\Ms = 0.6~\Msun$, the evolution is similar to the evolution for $1~\Msun$, with a maximum of $\delta P$ occurring at the transition saturated/unsaturated.
For $\Ms = 1.2~\Msun$, the star is always in the unsaturated regime so the shape is less complex. 
The wind is here acting to damp the effect of the planet even before the migration has finished.

As explained in \citet{Bolmont16}, due to the different values of the stellar dissipation and its evolution, the migration occurs faster for the more massive star and on longer timescales for the less massive star.
This means that the maximum of $\delta P$ occurs earlier for the more massive star and later for the less massive star.

\begin{figure*}[htbp!]
	\centering
%        \begin{center}
        \includegraphics[width=\linewidth]{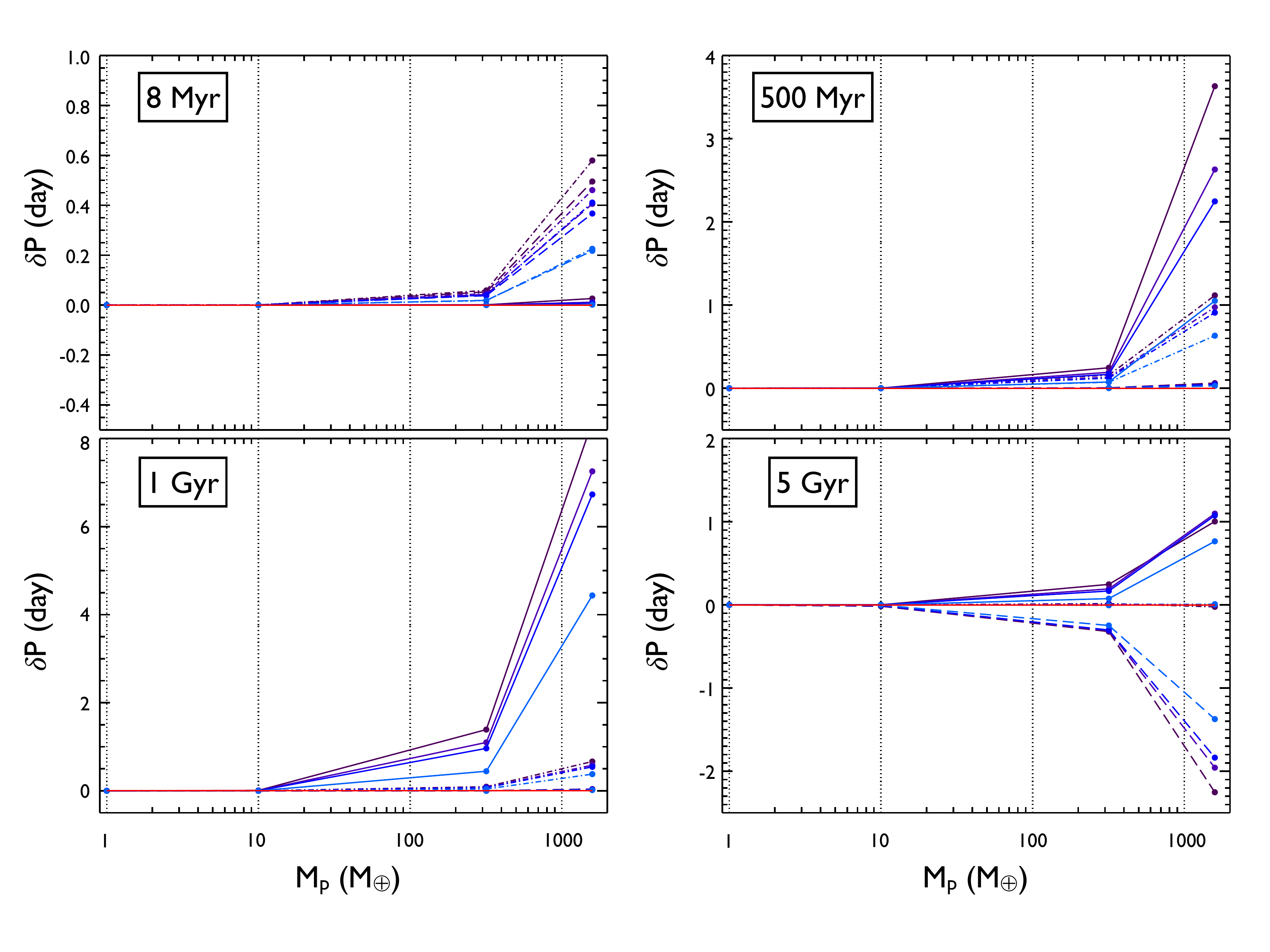}
        \caption{$\delta P$ for different initial orbital distances and for the 3 different stars and 4 different ages. The solid lines correspond to the $0.6~\Msun$ star, the dashed-dotted lines correspond to $1.0~\Msun$ and the long-dashed lines correspond to $1.2~\Msun$. From purple to light blue: an initial semi-major axis of 0.024, 0.028, 0.030 and 0.040~AU. The red line marks $\delta P = 0$~day. Note that the scales are different for each panel.} 
        \label{dP_age_Mp}
%        \end{center}
        \end{figure*}

In order to represent the evolution of $\delta P$ with the age of the star, we updated in Figure \ref{dP_age_Mp} the Figure 5 of \citet{Bolmont16} for four different ages: at 8~Myr (around the maximum effect for the star of $1.2~\Msun$), at 500~Myr (around the maximum effect for the star of $1.0~\Msun$), at 1~Gyr (around the maximum effect for the star of $0.6~\Msun$) and at 5~Gyr (the age used for Fig.~5 of \citealt{Bolmont16}).
We consider planet masses from the mass of Earth to five times the mass of Jupiter.

The effect of the planet on the star is much less pronounced than what was shown on Figure 5 of \citet{Bolmont16}.
The effect of a Jupiter mass planet on the rotation of a $0.6~\Msun$ star is of 0.1 to 0.3 days at 5~Gyr, while \citet{Bolmont16} was showing an effect of 2.2 to 3 days.
Note also that for the $1.2~\Msun$ star, the stars with planets rotate faster than the star without planets.
This coincides with the beginning of the red giant branch when the radius of the star starts to increase (see Fig.~\ref{fig:12}) and the planets have already started to migrate inwards. 

However, the planet can still have quite an important impact on the rotation towards the end of the outward migration.
This is thus an effect which highly depends on the age of the star, and which could eventually be observable via high-precision photometry \citep{McQuillan2013, Garcia2014, Ceillieretal2016}.

%%%%%%%%%%%%%%%%%%%%%%%%%%%%%%%%%%%%%%%%%%%%%%%%%%%%%%%%%%%%%%%

\begin{acknowledgements}  
The authors would like to thank the anonymous referee for his comments on this manuscript.
E. B. and S. M. acknowledge funding by the European Research Council through ERC grant SPIRE 647383.
\end{acknowledgements}

%\bibliographystyle{spmpsci}
%%\bibliographystyle{spbasic}
%
%%\bibliography{aamnem99,biblio}
%\bibliography{biblio}

\begin{thebibliography}{91}
\expandafter\ifx\csname natexlab\endcsname\relax\def\natexlab#1{#1}\fi

\bibitem[{{Bolmont} \& {Mathis}(2016)}]{Bolmont16}
{Bolmont}, E. \& {Mathis}, S. 2016, Celestial Mechanics and Dynamical
  Astronomy, 126, 275


%\bibitem[{{Albrecht} {et~al.}(2012){Albrecht}, {Winn}, {Johnson}, {Howard},
%  {Marcy}, {Butler}, {Arriagada}, {Crane}, {Shectman}, {Thompson}, {Hirano},
%  {Bakos}, \& {Hartman}}]{Albrechtetal2012}
%{Albrecht}, S., {Winn}, J.~N., {Johnson}, J.~A., {et~al.} 2012, ApJ, 757, 18
%
%\bibitem[{{Alexander}(1973)}]{Alexander1973}
%{Alexander}, M.~E. 1973, Ap\&SS, 23, 459
%
%\bibitem[{{Auclair-Desrotour} {et~al.}(2014){Auclair-Desrotour}, {Le
%  Poncin-Lafitte}, \& {Mathis}}]{AuclairDesrotouretal2014}
%{Auclair-Desrotour}, P., {Le Poncin-Lafitte}, C., \& {Mathis}, S. 2014, A\&A,
%  561, L7
%
%\bibitem[{{Auclair Desrotour} {et~al.}(2015){Auclair Desrotour}, {Mathis}, \&
%  {Le Poncin-Lafitte}}]{AuclairDesrotouretal2015}
%{Auclair Desrotour}, P., {Mathis}, S., \& {Le Poncin-Lafitte}, C. 2015, A\&A,
%  581, A118
%  
% \bibitem[{{Amard} {et~al.}(2016){Amard},{Palacios},{Charbonnel},{Gallet},{Bouvier}}]{Amardetal2016}
% {Amard}, L., {Palacios}, A., {Charbonnel}, C., {Gallet}, F., {Bouvier}, J. 2016, A\&A, 587, 105
%
%\bibitem[{{Baglin} {et~al.}(2006){Baglin}, {Auvergne}, {Boisnard}, {Lam-Trong},
%  {Barge}, {Catala}, {Deleuil}, {Michel}, \& {Weiss}}]{Baglin2006}
%{Baglin}, A., {Auvergne}, M., {Boisnard}, L., {et~al.} 2006, in COSPAR Meeting,
%  Vol.~36, 36th COSPAR Scientific Assembly
%
%\bibitem[{{Barker}(2011)}]{Barker2011}
%{Barker}, A.~J. 2011, MNRAS, 414, 1365
%
%\bibitem[{{Barker} \& {Lithwick}(2014)}]{Barkeretal2014}
%{Barker}, A.~J. \& {Lithwick}, Y. 2014, MNRAS, 437, 305
%
%\bibitem[{{Barker} \& {Ogilvie}(2009)}]{BO2009}
%{Barker}, A.~J. \& {Ogilvie}, G.~I. 2009, MNRAS, 395, 2268
%
%\bibitem[{{Barker} \& {Ogilvie}(2010)}]{BarkerOgilvie2010}
%{Barker}, A.~J. \& {Ogilvie}, G.~I. 2010, MNRAS, 404, 1849
%
%\bibitem[{{Barnes}(2003)}]{Barnes2003}
%{Barnes}, S.~A. 2003, ApJ, 586, 464
%
%\bibitem[{{Barnes} \& {Kim}(2010)}]{BarnesKim2010}
%{Barnes}, S.~A. \& {Kim}, Y.-C. 2010, ApJ, 721, 675
%
%\bibitem[{{Baruteau} \& {Rieutord}(2013)}]{BaruteauRieutord2013}
%{Baruteau}, C. \& {Rieutord}, M. 2013, Journal of Fluid Mechanics, 719, 47
%
%\bibitem[{{Bolmont} {et~al.}(2011){Bolmont}, {Raymond}, \&
%  {Leconte}}]{Bolmont2011}
%{Bolmont}, E., {Raymond}, S.~N., \& {Leconte}, J. 2011, A\&A, 535, A94
%
%\bibitem[{{Bolmont} {et~al.}(2015){Bolmont}, {Raymond}, {Leconte}, {Hersant},
%  \& {Correia}}]{Bolmont2015}
%{Bolmont}, E., {Raymond}, S.~N., {Leconte}, J., {Hersant}, F., \& {Correia},
%  A.~C.~M. 2015, A\&A, 583, A116
%
%\bibitem[{{Bolmont} {et~al.}(2012){Bolmont}, {Raymond}, {Leconte}, \&
%  {Matt}}]{Bolmont2012}
%{Bolmont}, E., {Raymond}, S.~N., {Leconte}, J., \& {Matt}, S.~P. 2012, A\&A,
%  544, A124
%
%\bibitem[{{Bonfils} {et~al.}(2013){Bonfils}, {Delfosse}, {Udry}, {Forveille},
%  {Mayor}, {Perrier}, {Bouchy}, {Gillon}, {Lovis}, {Pepe}, {Queloz}, {Santos},
%  {S{\'e}gransan}, \& {Bertaux}}]{Bonfils2013}
%{Bonfils}, X., {Delfosse}, X., {Udry}, S., {et~al.} 2013, A\&A, 549, A109
%
%\bibitem[{{Borucki} {et~al.}(2010){Borucki}, {Koch}, {Basri}, {Batalha},
%  {Brown}, {Caldwell}, {Caldwell}, {Christensen-Dalsgaard}, {Cochran},
%  {DeVore}, {Dunham}, {Dupree}, {Gautier}, {Geary}, {Gilliland}, {Gould},
%  {Howell}, {Jenkins}, {Kondo}, {Latham}, {Marcy}, {Meibom}, {Kjeldsen},
%  {Lissauer}, {Monet}, {Morrison}, {Sasselov}, {Tarter}, {Boss}, {Brownlee},
%  {Owen}, {Buzasi}, {Charbonneau}, {Doyle}, {Fortney}, {Ford}, {Holman},
%  {Seager}, {Steffen}, {Welsh}, {Rowe}, {Anderson}, {Buchhave}, {Ciardi},
%  {Walkowicz}, {Sherry}, {Horch}, {Isaacson}, {Everett}, {Fischer}, {Torres},
%  {Johnson}, {Endl}, {MacQueen}, {Bryson}, {Dotson}, {Haas}, {Kolodziejczak},
%  {Van Cleve}, {Chandrasekaran}, {Twicken}, {Quintana}, {Clarke}, {Allen},
%  {Li}, {Wu}, {Tenenbaum}, {Verner}, {Bruhweiler}, {Barnes}, \&
%  {Prsa}}]{Borucki2010}
%{Borucki}, W.~J., {Koch}, D., {Basri}, G., {et~al.} 2010, Science, 327, 977
%
%\bibitem[{{Bouvier}(2008)}]{Bouvier2008}
%{Bouvier}, J. 2008, A\&A, 489, L53
%
\bibitem[{{Bouvier} {et~al.}(1997){Bouvier}, {Forestini}, \&
  {Allain}}]{Bouvier1997}
{Bouvier}, J., {Forestini}, M., \& {Allain}, S. 1997, A\&A, 326, 1023
%
%\bibitem[{{Brun}(2014)}]{Brun2014}
%{Brun}, A.-S. 2014, Magnetic Fields throughout Stellar Evolution, Proceedings of the International Astronomical Union, IAU Symposium, 302, 114
%
\bibitem[{{Ceillier} {et~al.}(2016){Ceillier}, {van Saders}, {Garc{\'{\i}}a},
  {Metcalfe}, {Creevey}, {Mathis}, {Mathur}, {Pinsonneault}, {Salabert}, \&
  {Tayar}}]{Ceillieretal2016}
{Ceillier}, T., {van Saders}, J., {Garc{\'{\i}}a}, R.~A., {et~al.} 2016,
  MNRAS, 456, 119
%
%\bibitem[{{Charbonneau} {et~al.}(2000){Charbonneau}, {Brown}, {Latham}, \&
%  {Mayor}}]{Charbonneau2000}
%{Charbonneau}, D., {Brown}, T.~M., {Latham}, D.~W., \& {Mayor}, M. 2000, ApJ,
%  529, L45
%  
%\bibitem[{{Charbonneau}(2014)}]{Charbonneau2014}
%{Charbonneau}, P. 2014, Annual Review of Astronomy and Astrophysics, 52, 251
%
%\bibitem[{{Choi} \& {Herbst}(1996)}]{ChoiHerbst1996}
%{Choi}, P.~I. \& {Herbst}, W. 1996, The Astronomical Journal, 111, 283
%
%\bibitem[{{Correia} {et~al.}(2014){Correia}, {Bou{\'e}}, {Laskar}, \&
%  {Rodr{\'{\i}}guez}}]{Correiaetal2014}
%{Correia}, A.~C.~M., {Bou{\'e}}, G., {Laskar}, J., \& {Rodr{\'{\i}}guez}, A.
%  2014, A\&A, 571, A50
%
%\bibitem[{{Correia} \& {Laskar}(2003)}]{Correiaetal2003}
%{Correia}, A.~C.~M. \& {Laskar}, J. 2003, Journal of Geophysical Research
%  (Planets), 108, 9
%
%\bibitem[{{Damiani} \& {Lanza}(2015)}]{DamianiLanza2015}
%{Damiani}, C. \& {Lanza}, A.~F. 2015, A\&A, 574, A39
%
%\bibitem[{{Edwards} {et~al.}(1993){Edwards}, {Strom}, {Hartigan}, {Strom},
%  {Hillenbrand}, {Herbst}, {Attridge}, {Merrill}, {Probst}, \&
%  {Gatley}}]{Edwards1993}
%{Edwards}, S., {Strom}, S.~E., {Hartigan}, P., {et~al.} 1993, The Astronomical
%  Journal, 106, 372
%
%\bibitem[{{Efroimsky}(2012)}]{Efroimsky2012}
%{Efroimsky}, M. 2012, ApJ, 746, 150
%
%\bibitem[{{Efroimsky} \& {Makarov}(2013)}]{EM2013}
%{Efroimsky}, M. \& {Makarov}, V.~V. 2013, ApJ, 764, 26
%
%\bibitem[{{Eggleton} {et~al.}(1998){Eggleton}, {Kiseleva}, \& {Hut}}]{EKH1998}
%{Eggleton}, P.~P., {Kiseleva}, L.~G., \& {Hut}, P. 1998, ApJ, 499, 853
%
%\bibitem[{{Fabrycky} {et~al.}(2014){Fabrycky}, {Lissauer}, {Ragozzine}, {Rowe},
%  {Steffen}, {Agol}, {Barclay}, {Batalha}, {Borucki}, {Ciardi}, {Ford},
%  {Gautier}, {Geary}, {Holman}, {Jenkins}, {Li}, {Morehead}, {Morris},
%  {Shporer}, {Smith}, {Still}, \& {Van Cleve}}]{Fabryckyetal2014}
%{Fabrycky}, D.~C., {Lissauer}, J.~J., {Ragozzine}, D., {et~al.} 2014, ApJ,
%  790, 146
%
%\bibitem[{{Fang} \& {Margot}(2012)}]{FangMargot2012}
%{Fang}, J. \& {Margot}, J.-L. 2012, ApJ, 761, 92
%
%\bibitem[{{Favier} {et~al.}(2014){Favier}, {Barker}, {Baruteau}, \&
%  {Ogilvie}}]{Favieretal2014}
%{Favier}, B., {Barker}, A.~J., {Baruteau}, C., \& {Ogilvie}, G.~I. 2014,
%  MNRAS, 439, 845
%
%\bibitem[{{Ferraz-Mello} {et~al.}(2015){Ferraz-Mello}, {Tadeu dos Santos},
%  {Folonier}, {Czismadia}, {do Nascimento}, \&
%  {P{\"a}tzold}}]{Ferraz-Mello2015}
%{Ferraz-Mello}, S., {Tadeu dos Santos}, M., {Folonier}, H., {et~al.} 2015,
%  ApJ, 807, 78
%
%\bibitem[{{Gallet} \& {Bouvier}(2013)}]{GalletBouvier2013}
%{Gallet}, F. \& {Bouvier}, J. 2013, A\&A, 556, A36
%
%\bibitem[{{Gallet} \& {Bouvier}(2015)}]{GalletBouvier2015}
%{Gallet}, F. \& {Bouvier}, J. 2015, A\&A, 577, A98
%
\bibitem[{{Garc\'ia} {et~al.}(2014){Garc\'ia}, {Ceillier}, {Salabert},
  {Mathur}, {van Saders}, {Pinsonneault}, {Ballot}, {Beck}, {Bloemen},
  {Campante}, {Davies}, {do Nascimento}, {Mathis}, {Metcalfe}, {Nielsen},
  {Su{\'a}rez}, {Chaplin}, {Jim{\'e}nez}, \& {Karoff}}]{Garcia2014}
{Garc\'ia}, R.~A., {Ceillier}, T., {Salabert}, D., {et~al.} 2014, A\&A, 572,
  A34
%
%\bibitem[{{Gizon} {et~al.}(2013){Gizon}, {Ballot}, {Michel}, {Stahn},
%  {Vauclair}, {Bruntt}, {Quirion}, {Benomar}, {Vauclair}, {Appourchaux},
%  {Auvergne}, {Baglin}, {Barban}, {Baudin}, {Bazot}, {Campante}, {Catala},
%  {Chaplin}, {Creevey}, {Deheuvels}, {Dolez}, {Elsworth}, {Garcia}, {Gaulme},
%  {Mathis}, {Mathur}, {Mosser}, {Regulo}, {Roxburgh}, {Salabert}, {Samadi},
%  {Sato}, {Verner}, {Hanasoge}, \& {Sreenivasan}}]{Gizonetal2013}
%{Gizon}, L., {Ballot}, J., {Michel}, E., {et~al.} 2013, Proceedings of the
%  National Academy of Science, 110, 13267
%
%\bibitem[{{Goodman} \& {Dickson}(1998)}]{GD1998}
%{Goodman}, J. \& {Dickson}, E.~S. 1998, ApJ, 507, 938
%
%\bibitem[{{Goodman} \& {Lackner}(2009)}]{GL2009}
%{Goodman}, J. \& {Lackner}, C. 2009, ApJ, 696, 2054
%
%%\bibitem[{{Guenel} {et~al.}(2016){Guenel}, {Baruteau}, {Mathis}, \&
%%  {Rieutord}}]{Gueneletal2016}
%%{Guenel}, M., {Baruteau}, C., {Mathis}, S., \& {Rieutord}, M. 2016, ArXiv
%%  e-prints
%  \bibitem[Guenel et al.(2016)]{Gueneletal2016} Guenel, M., Baruteau, C., Mathis, S., Rieutord, M.\ 2016,\ A\&A, 589, A22
%
%\bibitem[{{Guenel} {et~al.}(2014){Guenel}, {Mathis}, \&
%  {Remus}}]{Gueneletal2014}
%{Guenel}, M., {Mathis}, S., \& {Remus}, F. 2014, A\&A, 566, L9
%
%\bibitem[{{Guillot} {et~al.}(2014){Guillot}, {Lin}, {Morel}, {Havel}, \&
%  {Parmentier}}]{Guillotetal2014}
%{Guillot}, T., {Lin}, D.~N.~C., {Morel}, P., {Havel}, M., \& {Parmentier}, V.
%  2014, in EAS Publications Series, Vol.~65, EAS Publications Series, 327--336
%%  \bibitem[Guillot et al.(2014)]{2014EAS....65..327G} Guillot, T., Lin, D.~N.~C., Morel, P., Havel, M., Parmentier, V.\ 2014.\ Evolution of exoplanets and their parent stars.\ EAS Publications Series 65, 327-336.
%
%\bibitem[{{Hansen}(2010)}]{Hansen2010}
%{Hansen}, B.~M.~S. 2010, ApJ, 723, 285
%
%\bibitem[{{Hansen}(2012)}]{Hansen2012}
%{Hansen}, B.~M.~S. 2012, ApJ, 757, 6
%
%\bibitem[{{Henning} {et~al.}(2009){Henning}, {O'Connell}, \&
%  {Sasselov}}]{Henningetal2009}
%{Henning}, W.~G., {O'Connell}, R.~J., \& {Sasselov}, D.~D. 2009, ApJ, 707,
%  1000
%
%\bibitem[{{Henry} {et~al.}(2000){Henry}, {Marcy}, {Butler}, \&
%  {Vogt}}]{Henry2000}
%{Henry}, G.~W., {Marcy}, G.~W., {Butler}, R.~P., \& {Vogt}, S.~S. 2000, ApJ,
%  529, L41
%
%\bibitem[{{Husnoo} {et~al.}(2012){Husnoo}, {Pont}, {Mazeh}, {Fabrycky},
%  {H{\'e}brard}, {Bouchy}, \& {Shporer}}]{Husnooetal2012}
%{Husnoo}, N., {Pont}, F., {Mazeh}, T., {et~al.} 2012, MNRAS, 422, 3151
%
%\bibitem[{{Hut}(1981)}]{Hut1981}
%{Hut}, P. 1981, A\&A, 99, 126
%
%\bibitem[{{Irwin} {et~al.}(2011){Irwin}, {Berta}, {Burke}, {Charbonneau},
%  {Nutzman}, {West}, \& {Falco}}]{Irwin2011}
%{Irwin}, J., {Berta}, Z.~K., {Burke}, C.~J., {et~al.} 2011, ApJ, 727, 56
%
%\bibitem[{{Ivanov} {et~al.}(2013){Ivanov}, {Papaloizou}, \&
%  {Chernov}}]{Ivanovetal2013}
%{Ivanov}, P.~B., {Papaloizou}, J.~C.~B., \& {Chernov}, S.~V. 2013, MNRAS, 432,
%  2339
%
%\bibitem[{{Jackson} {et~al.}(2008){Jackson}, {Greenberg}, \&
%  {Barnes}}]{Jackson2008}
%{Jackson}, B., {Greenberg}, R., \& {Barnes}, R. 2008, ApJ, 681, 1631
%
%\bibitem[{{Kaula}(1964)}]{Kaula1964}
%{Kaula}, W.~M. 1964, Reviews of Geophysics and Space Physics, 2, 661
%
%\bibitem[{{Kawaler}(1988)}]{Kawaler1988}
%{Kawaler}, S.~D. 1988, ApJ, 333, 236
%
%\bibitem[{{Lai}(2012)}]{Lai2012}
%{Lai}, D. 2012, MNRAS, 423, 486
%
%\bibitem[{{Lanza}(2010)}]{Lanza2010}
%{Lanza}, A.~F. 2010, A\&A, 512, A77
%
%\bibitem[{{Lanza} \& {Shkolnik}(2014)}]{LanzaShkolnik2014}
%{Lanza}, A.~F. \& {Shkolnik}, E.~L. 2014, MNRAS, 443, 1451
%
%\bibitem[{{Leconte} {et~al.}(2010){Leconte}, {Chabrier}, {Baraffe}, \&
%  {Levrard}}]{Leconte2010}
%{Leconte}, J., {Chabrier}, G., {Baraffe}, I., \& {Levrard}, B. 2010, A\&A,
%  516, A64+
%
%\bibitem[{{MacGregor} \& {Brenner}(1991)}]{MacGregorBrenner1991}
%{MacGregor}, K.~B. \& {Brenner}, M. 1991, ApJ, 376, 204
%
%%\bibitem[{{Maeder}(2009)}]{Maeder2009}
%%{Maeder}, A. 2009, {Physics, Formation and Evolution of Rotating Stars}
\bibitem[\protect\citeauthoryear{Maeder}{2009}]{2009pfer.book.....M} Maeder, A.\ 2009.\ Physics, Formation and Evolution of Rotating Stars.\, Springer Berlin Heidelberg.

%\bibitem[\protect\citeauthoryear{Maeder}{2009}]{2009pfer.book.....M} Maeder A., 2009, pfer.book

%
%\bibitem[{{Mathis}(2015{\natexlab{a}})}]{Mathis2015a}
%{Mathis}, S. 2015{\natexlab{a}}, in SF2A-2015: Proceedings of the Annual
%  meeting of the French Society of Astronomy and Astrophysics. Eds.: F.
%  Martins, S. Boissier, V. Buat, L. Cambr{\'e}sy, P. Petit, pp.401-405, ed.
%  F.~{Martins}, S.~{Boissier}, V.~{Buat}, L.~{Cambr{\'e}sy}, \& P.~{Petit},
%  401--405
%
%\bibitem[{{Mathis}(2015{\natexlab{b}})}]{Mathis2015b}
%{Mathis}, S. 2015{\natexlab{b}}, A\&A, 580, L3
%
%\bibitem[{{Mathis} \& {Zahn}(2004)}]{MZ2004}
%{Mathis}, S. \& {Zahn}, J.-P. 2004, A\&A, 425, 229
%
%\bibitem[{{Mathis} \& {Le Poncin-Lafitte}(2009)}]{MLP09}
%{Mathis}, S. \& {Le Poncin-Lafitte}, C. 2009, A\&A, 497, 889
%
%\bibitem[{{Mathis} \& {Remus}(2013)}]{MR2013}
%{Mathis}, S. \& {Remus}, F. 2013, Lecture Notes in Physics (The Environments of the Sun and the Stars, ed. by J.-P. Rozelot and C. Neiner), Volume 857, 111
%
%\bibitem[{{Matt} {et~al.}(2015){Matt}, {Brun}, {Baraffe}, {Bouvier}, \&
%  {Chabrier}}]{Matt2015}
%{Matt}, S.~P., {Brun}, A.~S., {Baraffe}, I., {Bouvier}, J., \& {Chabrier}, G.
%  2015, ApJ, 799, L23
%
%\bibitem[{{Mayor} \& {Queloz}(1995)}]{MayorQueloz1995}
%{Mayor}, M. \& {Queloz}, D. 1995, Nature, 378, 355
%
\bibitem[{{McQuillan} {et~al.}(2013){McQuillan}, {Mazeh}, \&
  {Aigrain}}]{McQuillan2013}
{McQuillan}, A., {Mazeh}, T., \& {Aigrain}, S. 2013, ApJ, 775, L11
%
%\bibitem[{{McQuillan} {et~al.}(2014){McQuillan}, {Mazeh}, \&
%  {Aigrain}}]{McQuillanetal2014}
%{McQuillan}, A., {Mazeh}, T., \& {Aigrain}, S. 2014, ApJS, 211, 24
%
%\bibitem[{{Mignard}(1979)}]{Mignard1979}
%{Mignard}, F. 1979, Moon and Planets, 20, 301
%
%\bibitem[{{Ogilvie}(2013)}]{Ogilvie2013}
%{Ogilvie}, G.~I. 2013, MNRAS, 429, 613
%
%\bibitem[{{Ogilvie}(2014)}]{Ogilvie2014}
%{Ogilvie}, G.~I. 2014, ARA\&A, 52, 171
%
%\bibitem[{{Ogilvie} \& {Lin}(2004)}]{OgilvieLin2004}
%{Ogilvie}, G.~I. \& {Lin}, D.~N.~C. 2004, ApJ, 610, 477
%
%\bibitem[{{Ogilvie} \& {Lin}(2007)}]{OgilvieLin2007}
%{Ogilvie}, G.~I. \& {Lin}, D.~N.~C. 2007, ApJ, 661, 1180
%
%\bibitem[{{Paz-Chinch{\'o}n} {et~al.}(2015){Paz-Chinch{\'o}n}, {Le{\~a}o},
%  {Bravo}, {de Freitas}, {Ferreira Lopes}, {Alves}, {Catelan}, {Canto Martins},
%  \& {De Medeiros}}]{Paz2015}
%{Paz-Chinch{\'o}n}, F., {Le{\~a}o}, I.~C., {Bravo}, J.~P., {et~al.} 2015, ApJ,
%  803, 69
%
%\bibitem[{{Penev} {et~al.}(2014){Penev}, {Zhang}, \& {Jackson}}]{Penevetal2014}
%{Penev}, K., {Zhang}, M., \& {Jackson}, B. 2014, PASP, 126, 553
%
%%\bibitem[{{Perryman}(2011)}]{Perryman2011}
%%{Perryman}, M. 2011, {The Exoplanet Handbook}
%\bibitem[Perryman(2011)]{Perryman2011} Perryman, M.\ 2011.\ The Exoplanet Handbook.\, Cambridge University Press. 
%
%\bibitem[{{Pont}(2009)}]{Pont2009}
%{Pont}, F. 2009, MNRAS, 396, 1789
%
%\bibitem[{{Poppenhaeger} \& {Wolk}(2014)}]{Pop2014}
%{Poppenhaeger}, K. \& {Wolk}, S.~J. 2014, A\&A, 565, L1
%
%\bibitem[{{Rebull} {et~al.}(2006){Rebull}, {Stauffer}, {Megeath}, {Hora}, \&
%  {Hartmann}}]{Rebull2006}
%{Rebull}, L.~M., {Stauffer}, J.~R., {Megeath}, S.~T., {Hora}, J.~L., \&
%  {Hartmann}, L. 2006, ApJ, 646, 297
%
%\bibitem[{{Rebull} {et~al.}(2004){Rebull}, {Wolff}, \& {Strom}}]{Rebull2004}
%{Rebull}, L.~M., {Wolff}, S.~C., \& {Strom}, S.~E. 2004, The Astronomical
%  Journal, 127, 1029
%
%\bibitem[{{Remus} {et~al.}(2012{\natexlab{a}}){Remus}, {Mathis}, \&
%  {Zahn}}]{RMZ2012}
%{Remus}, F., {Mathis}, S., \& {Zahn}, J.-P. 2012{\natexlab{a}}, A\&A, 544, A132
%
%\bibitem[{{Remus} {et~al.}(2012{\natexlab{b}}){Remus}, {Mathis}, {Zahn}, \&
%  {Lainey}}]{RMZL2012}
%{Remus}, F., {Mathis}, S., {Zahn}, J.-P., \& {Lainey}, V. 2012{\natexlab{b}},
%  A\&A, 541, A165
%
%\bibitem[{{R{\'e}ville} {et~al.}(2015){R{\'e}ville}, {Brun}, {Matt},
%  {Strugarek}, \& {Pinto}}]{Reville2015}
%{R{\'e}ville}, V., {Brun}, A.~S., {Matt}, S.~P., {Strugarek}, A., \& {Pinto},
%  R.~F. 2015, ApJ, 798, 116
%
%\bibitem[{{Savonije}(2008)}]{S2008}
%{Savonije}, G.-J. 2008, in EAS Publications Series, Vol.~29, EAS Publications
%  Series, ed. M.-J. {Goupil} \& J.-P. {Zahn}, 91--125
%
\bibitem[{{Siess} {et~al.}(2000){Siess}, {Dufour}, \& {Forestini}}]{Siess2000}
{Siess}, L., {Dufour}, E., \& {Forestini}, M. 2000, A\&A, 358, 593
%
%\bibitem[{{Skumanich}(1972)}]{Skumanich1972}
%{Skumanich}, A. 1972, ApJ, 171, 565
%
%\bibitem[{{Teitler} \& {K{\"o}nigl}(2014)}]{Teitler2014}
%{Teitler}, S. \& {K{\"o}nigl}, A. 2014, ApJ, 786, 139
%
%\bibitem[{{Terquem} {et~al.}(1998){Terquem}, {Papaloizou}, {Nelson}, \&
%  {Lin}}]{Terquemetal1998}
%{Terquem}, C., {Papaloizou}, J.~C.~B., {Nelson}, R.~P., \& {Lin}, D.~N.~C.
%  1998, ApJ, 502, 788
%
%\bibitem[{{Tobie} {et~al.}(2005){Tobie}, {Mocquet}, \& {Sotin}}]{Tobieetal2005}
%{Tobie}, G., {Mocquet}, A., \& {Sotin}, C. 2005, Icarus, 177, 534
%
%%\bibitem[{{van Saders} {et~al.}(2016){van Saders}, {Ceillier}, {Metcalfe},
%%  {Silva Aguirre}, {Pinsonneault}, {Garc{\'{\i}}a}, {Mathur}, \&
%%  {Davies}}]{VanSadersetal2016}
%%{van Saders}, J.~L., {Ceillier}, T., {Metcalfe}, T.~S., {et~al.} 2016, ArXiv
%%  e-prints
%  \bibitem[van Saders et al.(2016)]{VanSadersetal2016} van Saders, J.~L., Ceillier, T., Metcalfe, T.~S., Silva Aguirre, V., Pinsonneault, M.~H., Garc{\'{\i}}a, R.~A., Mathur, S., Davies, G.~R.\ 2016, Nature 529, 181-184
%
%\bibitem[{{Winn} {et~al.}(2010){Winn}, {Fabrycky}, {Albrecht}, \&
%  {Johnson}}]{Winnetal2010}
%{Winn}, J.~N., {Fabrycky}, D., {Albrecht}, S., \& {Johnson}, J.~A. 2010, ApJ,
%  718, L145
%
%\bibitem[{{Witte} \& {Savonije}(1999)}]{WS1999}
%{Witte}, M.~G. \& {Savonije}, G.~J. 1999, A\&A, 350, 129
%
%\bibitem[{{Zahn}(1966)}]{Zahn1966}
%{Zahn}, J.~P. 1966, Annales d'Astrophysique, 29, 489
%
%\bibitem[{{Zahn}(1975)}]{Zahn1975}
%{Zahn}, J.-P. 1975, A\&A, 41, 329
%
%\bibitem[{{Zahn}(1977)}]{Zahn1977}
%{Zahn}, J.-P. 1977, A\&A, 57, 383
%
%\bibitem[{{Zahn}(1989)}]{Zahn1989}
%{Zahn}, J.-P. 1989, A\&A, 220, 112
%
%\bibitem[{{Zahn} \& {Bouchet}(1989)}]{ZahnBouchet1989}
%{Zahn}, J.-P. \& {Bouchet}, L. 1989, A\&A, 223, 112
%
%\bibitem[{{Zahn}(1992)}]{Zahn1992}
%{Zahn}, J.-P. 1992, A\&A, 265, 115
%
%\bibitem[{{Zahn}(1994)}]{Zahn1994}
%{Zahn}, J.-P. 1994, A\&A, 288, 829

\end{thebibliography}


\end{document}